\documentclass[twocolumn,usenatbib]{aastex62}

\usepackage{colortbl}
\usepackage{multirow}
\usepackage{blindtext}
\usepackage{tabularx}
\usepackage{textcomp}

\received{\ldots}
\revised{\ldots}
\accepted{\ldots}
\published{\ldots}
\reportnum{Accepted on 9/6/2020}

\shorttitle{The slow heartbeats of a ULX in NGC\,3621}
\shortauthors{Motta et al.}

\begin{document}

\title{The slow heartbeats of an ultra-luminous X-ray source in NGC\,3621}


\correspondingauthor{S. E. Motta}
\email{sara.motta@physics.ox.ac.uk}
\author[0000-0002-6154-5843]{S.\,E. Motta}
 \affil{Department of Physics, Astrophysics, University of Oxford, Denys Wilkinson Building, Keble Road, OX1 3RH Oxford, UK}
\affil{INAF--Osservatorio Astronomico di Brera, via E.\,Bianchi 46, 23807 Merate (LC), Italy}
\author[0000-0002-8017-0338]{M. Marelli}
\author[0000-0002-3869-2925]{F. Pintore}
\affil{INAF--Istituto di Astrofisica Spaziale e Fisica Cosmica di Milano, via A.\,Corti 12, 20133 Milano, Italy}
\author[0000-0003-4849-5092]{P. Esposito}
\affil{Scuola Universitaria Superiore IUSS Pavia, Palazzo del Broletto, piazza della Vittoria 15, 27100 Pavia, Italy}
\affil{INAF--Istituto di Astrofisica Spaziale e Fisica Cosmica di Milano, via A.\,Corti 12, 20133 Milano, Italy}
\author[0000-0002-9393-8078]{R. Salvaterra}
\affil{INAF--Istituto di Astrofisica Spaziale e Fisica Cosmica di Milano, via A.\,Corti 12, 20133 Milano, Italy}
\author[0000-0001-6739-687X]{A. De Luca}
\affil{INAF--Istituto di Astrofisica Spaziale e Fisica Cosmica di Milano, via A.\,Corti 12, 20133 Milano, Italy}
\affil{INFN, Sezione di Pavia, Via A. Bassi 6, 27100 Pavia, Italy}
\author[0000-0001-5480-6438]{G.\,L. Israel}
\affil{INAF--Osservatorio Astronomico di Roma, via Frascati 33, 00078 Monteporzio Catone, Italy}
\author[0000-0002-6038-1090]{A. Tiengo}
\affil{Scuola Universitaria Superiore IUSS Pavia, Palazzo del Broletto, piazza della Vittoria 15, 27100 Pavia, Italy}
\affil{INAF--Istituto di Astrofisica Spaziale e Fisica Cosmica di Milano, via A.\,Corti 12, 20133 Milano, Italy}
\affil{INFN, Sezione di Pavia, Via A. Bassi 6, 27100 Pavia, Italy}
\author[0000-0003-3952-7291]{G.\,A. Rodr\'iguez Castillo}
\affil{INAF--Osservatorio Astronomico di Roma, via Frascati 33, 00078 Monteporzio Catone, Italy}
\def\xmm {\emph{XMM--Newton}}
\def\nus {\emph{NuSTAR}}
\def\cxo {\emph{Chandra}}
\def\swift {\emph{Swift}}
\def\src {\mbox{4X\,J1118}}
\def\flux {\mbox{erg cm$^{-2}$ s$^{-1}$}}
\def\lum {\mbox{erg s$^{-1}$}}
\def\nh {$N_{\rm H}$}

\def\GRS {\mbox{GRS\,1915+105}}
\def\IGR {\mbox{IGR\,J17091--3624}}
\def\RB {\mbox{Rapid Burster}}
\def\GSN {\mbox{GSN\,069}}

\begin{abstract}

\noindent We report on the results of X-ray observations of 4XMM\,J111816.0--324910, a transient ultra-luminous X-ray source located in the galaxy NGC\,3621. 
This system is characterised by a transient nature and marked variability with characteristic time-scale of $\approx$3500\,s, differently from other ULXs, which in the vast majority show limited intra-observation variability. Such a behaviour is very reminiscent of the so-called \textit{heartbeats} sometimes observed in the Galactic black hole binary GRS\,1915+105, where the variability time-scale is $\sim$10--1000\,s. 
We study the spectral and timing properties of this object and find that overall, once the differences in the variability time-scales are taken into account, they match quite closely those of both GRS\,1915+105, and of a number of objects showing heartbeats in their light-curves, including a confirmed neutron star and a super-massive black hole powering an active galactic nucleus. 
We investigate the nature of the compact object in 4XMM\,J111816.0--324910 by searching for typical neutron star signatures and by attempting a mass estimate based on different methods and assumptions. Based on the current available data, we are not able to unambiguously determine the nature of the accreting compact object responsible for the observed phenomenology.
\end{abstract}

\keywords{Compact object : Black holes -- Compact object : Neutron stars -- Accretion -- X-ray point sources -- Extragalactic astronomy } 

\section{Introduction} \label{sec:intro}

The detection and characterization of X-ray variability play a role of paramount importance in the process of identifying the nature of a source and studying the mechanisms powering the observed emission. In accreting sources, it offers unique insights into the condition of disks, accretion flows, winds and jets in the extreme environments surrounding a compact object. 
In this work, we investigate a peculiar X-ray source in  the  galaxy  NGC\,3621, listed in the 4XMM catalogue (Webb et al., submitted) as 4XMM\,J111816.0--324910  (hereafter \src). This source was singled out due to its marked variability during a large study of the soft X-ray sky in the temporal domain, carried out via tools developed as part of the EXTraS project\footnote{\emph{Exploring the X-ray Transient and variable Sky} \citep{deluca16} was a project aimed at mining the \xmm\ archival data for periodic and aperiodic variability in the time domain at all time scales (limited to the 3XMM-DR5 catalog observations; \citealt{rosen16}). See \url{http://www.extras-fp7.eu}}, which were applied to all the XMM-Newton observations performed up to the end of 2018.

NGC\,3621 is a spiral galaxy located at a cepheid distance of 6.7\,Mpc \citep{tully13}.
Despite the bulgeless morphology of NGC\,3621, \citet{Satyapal2007} discovered a faint active galactic nucleus (AGN) in its nuclear cluster from mid-infrared observations and, based on estimates of its bolometric luminosity, set a lower limit on the mass of the central black hole (BH) of $4\times10^3\,M_\odot$. 
\citet{Barth2009} classified NGC\,3621 as a Seyfert 2 galaxy based on its optical spectrum, and from stellar-dynamical modeling of the nuclear cluster 
set an upper limit to the BH mass of $3\times10^6\,M_\odot$. 
Using a \cxo\ observation perfomed in March 2008, \citet{Gliozzi2009} found further evidence for an AGN in NGC\,3621 with the detection of a weak X-ray point source coincident with the nucleus. 

In the same observation, they also detected two bright off-nucleus sources [one  in  the ultra-luminous X-ray source  (ULX; \citealt{kaaret17}) regime] but did not focus on \src, which at the time was at the inconspicuous luminosity of $L_{\mathrm{X}}\approx10^{37}$\,\lum\ (see Sec.\,\ref{sect:otherdata}), yielding only a handful of photons. 
In a more recent \xmm\ observation (taken in December 2017), we  detected \src\ at more than $3\times10^{39}$\,\lum, which qualifies it as a new (transient) ULX.\footnote{We note that also the fainter non-nuclear source studied by \citet{Gliozzi2009}, their `source C', was above the ULX threshold in the same \xmm\ observation.} Even more surprisingly, the inspection of its light-curve revealed an unusual quasi-periodic modulation of the flux, very reminiscent of that seen at times in the Galactic BH X-ray binary GRS\,1915+105, the so-called \textit{heartbeat} oscillations (formally known as $\rho$-class variability, according to the classification proposed by \citealt{Belloni2000}).

In the following, we report on the timing and spectral characteristics of \src, based on the \xmm\ observation, and discuss the nature of this very peculiar source, also using data from \cxo, the \emph{Swift Neil Gehrels Observatory},  \nus, and \emph{HST}.

\section{XMM-Newton observation, data reduction and analysis}

The \xmm\ observation that showed \src \ as a ULX (see Table\,\ref{table:obs}) comprises data from both EPIC-pn and the EPIC-MOS1 and MOS2 CCD cameras. Because of the low flux and complicated field, the RGS did not provide useful data. All cameras were operated in \textit{Full-Frame} mode, which yields data with a time resolution of 73.4\,ms for the EPIC-pn,  2.7\,s for the MOS1, and 2.6\,s for the MOS2. The data reduction was performed following standard procedures using SAS v.16.1. We selected events with PATTERN$\leq$4 and PATTERN$\leq$12 for the pn and the MOS, respectively. 
Photons for both the timing and spectral analysis were extracted in a circular regions with radius of 15$''$ centered on \src\ (this small radius, $\sim$70\% enclosed energy fraction, is due to the presence of nearby sources), while the background was evaluated from a larger source-free region in the same chip as \src. In the following, we only considered events collected when the EPIC-pn and MOS cameras were observing the target at the same time (approximately 35\,ks).

\subsection{Timing}

\begin{figure}[ht]
\begin{center}
\hspace{-8mm}
\includegraphics[width=0.48\textwidth]{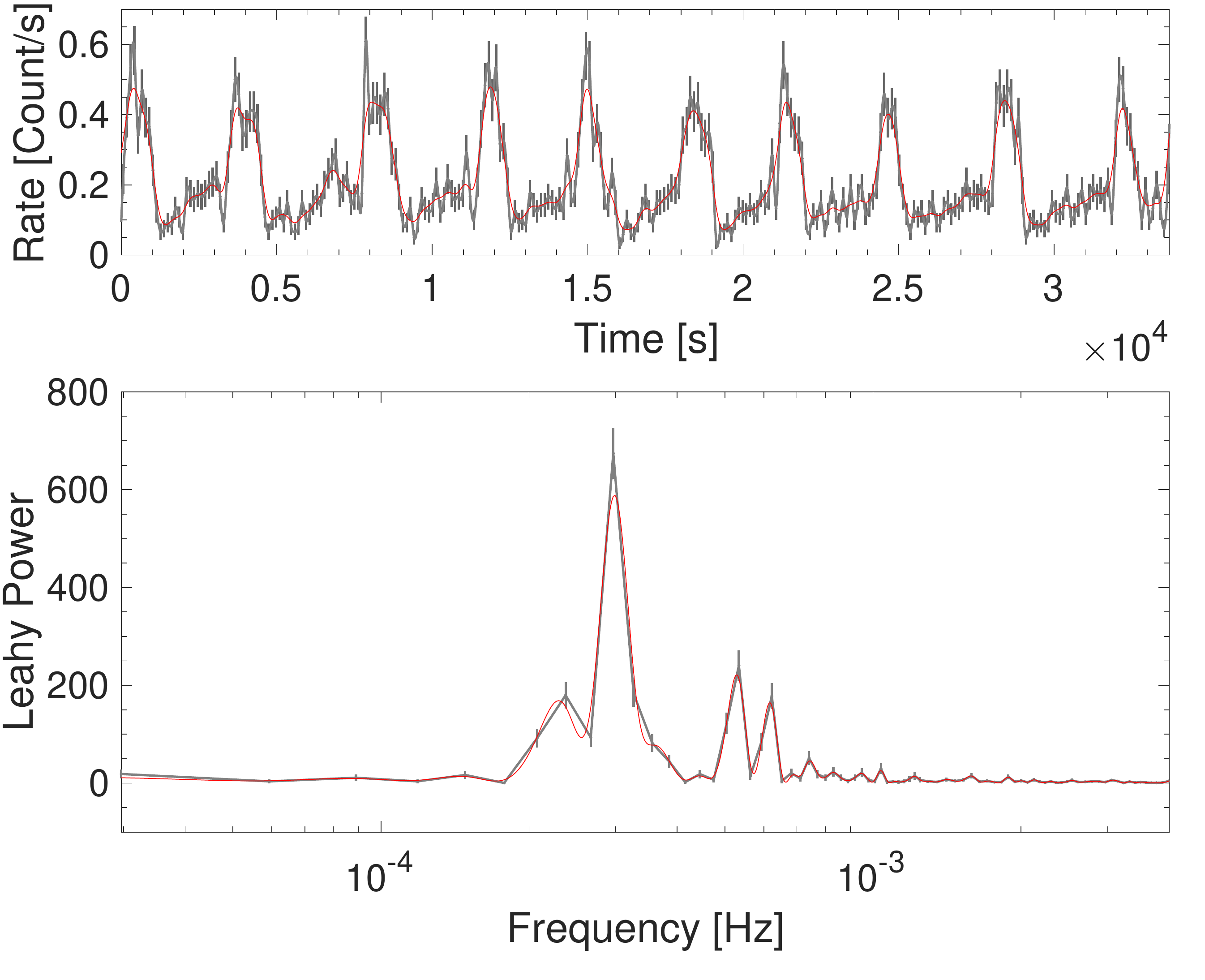}
\caption{\xmm\ data from \src. \textit{Top panel:} Light-curve extracted combining the events collected with all EPIC cameras. \textit{Bottom panel:} Power density spectrum calculated from the same data. In both panels, the actual data and their uncertainties (in grey) were smoothed using a Gaussian filter for clarity (red solid line).}
\end{center}
\label{fig:XMM_PDS}
\end{figure}

We created three light-curves of \src \ by combining events collected with EPIC-pn, EPIC-MOS1 and EPIC-MOS2 in the energy bands 0.3--10\,keV, 0.3--2\,keV, and 2--10\,keV. We then binned the data in three light-curves with a 120\,s time bin (Nyquist frequency $\rm{Ny} = 0.0083$\,Hz), which corresponds to our final time resolution.
In order to obtain the best possible frequency resolution ($\delta \nu = 1/T \approx 3\times 10^{-5}$\,Hz), we calculated one single power density spectrum (PDS) from the entire light-curve extracted in the 0.3--10\,keV energy band (duration $T_{\text{obs}} \approx 34$\,ks). The resulting light-curve and PDS are shown in Fig.\,\ref{fig:XMM_PDS} (top and bottom panel, respectively). 

\begin{figure}[ht]
\begin{center}
\hspace{-8mm}
\includegraphics[width=0.48\textwidth]{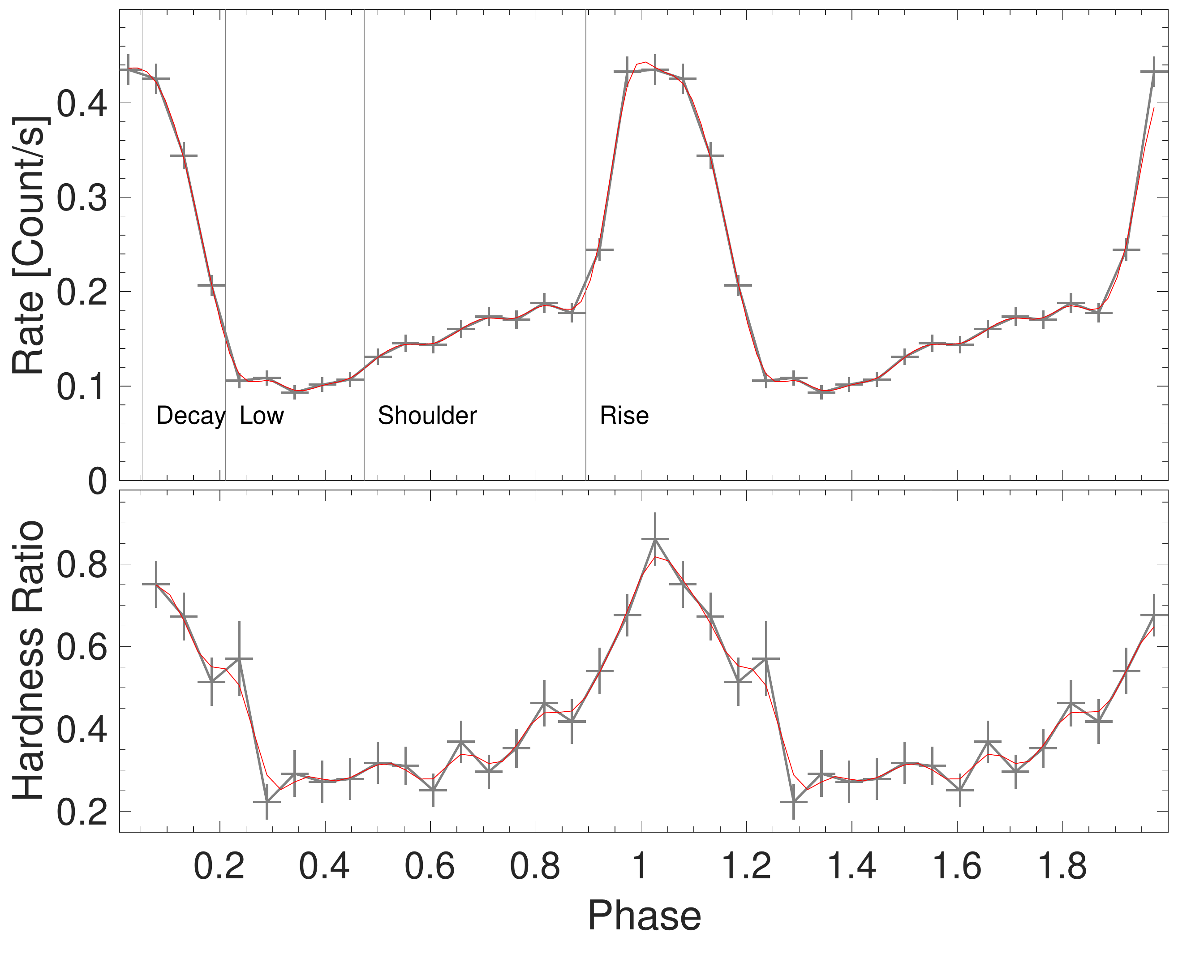}
\caption{\textit{Top panel}: EPIC phase-folded light-curve, in the 0.3--10\,keV band. The phase intervals used for the phase-resolved spectroscopy are marked by vertical lines. \textit{Bottom panel}: Hardness ratio between the fluxes obtained in the 0.3--2\,keV and 2--10\,keV bands, respectively. As in Fig. \ref{fig:XMM_PDS} the actual data were smoothed using a Gaussian filter for clarity.}
\end{center}
\label{fig:XMM_folded}
\end{figure}

The EPIC light-curve from \src\ clearly shows a repeating pattern, although not strictly periodic. Therefore, a simple folding of the light-curve would not return an accurate representation of the variability that we observe. On the other hand, the relatively low flux of this source requires some sort of averaging to allow a more accurate data analysis to be performed. 
We created a folded light-curve profile following the approach of \cite{Neilsen2011}, who applied it to GRS\,1915+105. 
We first constructed a first-guess template model defining a representative single light-curve cycle. We then cross-correlated the template with the entire light-curve, which comprises of 9 peaks. The resulting cross-correlation returns 9 maxima that correspond to the peak in the light-curve, and are assigned phase zero ($\phi=0$) in the construction of the folded profile. We then split the cross correlation function in 20 phase bins, over which we phase-average the original light-curve. We then use the resulting profile as a new template, which we cross-correlate with the original light-curve. The process is then iterated for 10 times so to minimise the bias introduced with the choice of our first-guess template. 
We applied the above method to the three light-curves we extracted, and we used the light-curves in the 0.3--2.0\,keV and 3.0--10\,keV bands to calculate a hardness ratio.   
The final phase-folded light-curve extracted for the 0.3--10\,keV band and the hardness ratio as a function of phase are shown in Fig. \ref{fig:XMM_folded} (top and bottom panel, respectively). The profiles shown  correspond to an average period of approximately 3500\,s. 

In the following, uncertainties will be given at 1$\sigma$, with the exception of the spectral results, where uncertainties are quoted at a 90\% confidence level.


\subsection{Spectral analysis}

We first extracted the three (one EPIC-pn and two EPIC-MOS) average spectra in the \mbox{0.3--10\,keV} energy range from the entire observation. The appropriate ancillary and response files were created with the SAS package. 
Then, in order to probe the fast variability of the target along the repeated pattern that characterises the light-curve, we extracted 4 spectra in the 0.3--10 keV energy range at different phases of said pattern, following the folded profile obtained as described above. We extracted 4 spectra, labelled \textit{low}, \textit{shoulder}, \textit{rise} and \textit{decay}, in the phase ranges shown in Fig.\,\ref{fig:XMM_folded} (top panel).

\section{Results}

\subsection{Timing results}

The PDS from \src \ shows a prominent main peak at ($3.02\pm 0.03$) $\times 10^{-4}$\,Hz, clearly corresponding to the obvious high-amplitude $\approx$3500\,s modulation visible in the light-curve. Such a peak is detected with high statistical significance (15.3$\sigma$) and has a quality factor $Q = 18$.\footnote{The quality factor $Q$ quantifies the coherence of a given signal, and is defined as $Q = \nu/\Delta \nu$, where $\nu$ and $\Delta \nu$ are the centroid frequency and the FWHM of the peak, respectively.}
The PDS shows two additional secondary peaks, at ($5.2\pm0.1$)$\times 10^{-4}$\,Hz and ($7\pm2$)$\times 10^{-4}$\,Hz (9.2 and 7.6$\sigma$, respectively), although they are both unresolved (i.e., their FWHM is significantly lower than our frequency resolution of $\sim 3\times 10^{-5}$\,s). A relatively low amplitude peak at ($2.22\pm0.01$)$\times 10^{-4}$\,Hz (7.2$\sigma$) is also detected. 
While the peak at 6$\times 10^{-4}$\,Hz is consistent with being harmonically related to the prominent main peak, both the peaks at 5.2 and 2.22$\times 10^{-4}$\,Hz do not appear to be in any obvious harmonic relation with the main peak, and cannot be obviously associated with a particular feature in the light-curve. Further in-depth analysis of the above features is beyond the scope of this work.

For the sake of comparison, we analysed a representative \emph{RXTE} observation of GRS\,1915+105 (Obs. ID 60405-01-02-00) during which the heartbeats were clearly observed. 
The flare modulation occurs on an average period of 50\,s, and---similarly to \src---is clearly visible as a prominent peak in the PDS. The PDS shows also two secondary peaks, qualitatively similar to those observed in the PDS from \src. In this case, however, both peaks are consistent with being harmonically related to the main peak, forming a 1-2-3 harmonic series.\footnote{Note that while third harmonics are not unheard of in accreting BHs binaries (see e.g. \citealt{Motta2017}), they remain puzzling and their origin widely debated.} We note that this is only a representative case, and that GRS\,1915+105 shows similar modulations at a variable period, which can be as long as $\sim$1000\,s \citep{Belloni2000, Altamirano2011, Weng2018}. Hence, the properties of the PDS change in time, and the harmonic relation highlighted above is not always observed nor consistently respected.   

\smallskip

The \src\ profile that emerges in its folded light-curve has a clear structure, formed by: a main peak, consistent with being symmetrical and best-described by a Gaussian function; a left shoulder, preceding the peak; a low flux phase, corresponding to the minimum flux level reached during the duty cycle, following the main peak.
Based on the above, we can estimate a number of time-scales that characterise the variability in \src. We stress that the repeated pattern in the light-curve is not strictly periodic, and  therefore the timescales in the following have to be intended as average quantities.   
The average \textit{recurrence time} of the flaring observed in the light-curve is $T_{\mathrm{rec}}\approx3500$\,s. The flares are consistent with being almost symmetric, and have a \textit{duration} of $T_{\mathrm{dur}}\approx 700$\,s, which is defined as the FWHM of the Gaussian that best fits the main peak of the folded light-curve. The average \textit{flare rise} and \textit{decay} are equal to $T_{\mathrm{rise}} \approx 350$\,s and $T_{\mathrm{decay}}\approx 525$\,s, respectively. We defined $T_{\mathrm{rise}}$ as the interval between the intersection of the Gaussian that best-fits the flare and the \textit{shoulder} preceding the flare itself, and the Gaussian peak. The decay is defined as the interval between the peak of said Gaussian profile and the first point of the plateau following the flare (around $\phi$=0.2).
We note that the $T_{\mathrm{rise}}$ and $T_{\mathrm{decay}}$ time-scales do not exactly coincide with the phase ranges we defined to perform the phase-resolved spectroscopy (where the limited signal-to-noise ratio imposed a less accurate event selection with respect to that adopted for the timing analysis).

\subsection{Spectral results}
The EPIC average energy spectra are well-fit by a number of models, which include a blackbody (bbody), a disk-blackbody (diskbb), and a power law or a bremsstrahlung profile (bremss), all modified by a component accounting for the absorption by the inter-stellar medium (tbabs, using the abundances by \citealt{Wilms00}). Unsurprisingly, combinations of the above models or similar ones (such as  diskpbb or Comptonization models, e.g., nthcomp; \citealt{Zdziarski1996}) also provide acceptable fits. 
For example, an absorbed diskbb yields a $\chi_\nu^2=1.13$ for 267 degrees of freedom (dof), with the following parameters: \mbox{$N_{\mathrm{H}}=(3.0\pm1.0)\times10^{20}$\,cm$^{-2}$}, \mbox{$kT=1.6\pm0.7$\,keV}, and
an observed \mbox{0.3--10\,keV} flux of \mbox{$(5.9\pm0.2)\times10^{-13}$\,\flux}.
For an absorbed diskpbb \mbox{($\chi^2/\mathrm{dof}=0.95/266$)} we obtained $N_{\mathrm{H}}=(1.3\pm0.3)\times10^{20}$\,cm$^{-2}$, \mbox{$kT=2.4_{-0.3}^{+0.5}$\,keV}, and \mbox{$p=0.59\pm0.02$}, providing an absorbed flux of $(6.2\pm0.3)\times10^{-13}$\,\flux. 
For a distance of 6.7\,Mpc, the inferred average luminosity is \mbox{$(3.24\pm0.08)\times10^{39}$} and \mbox{$(3.7\pm0.1)\times10^{39}$\,\lum} for the  diskbb and  diskpbb model, respectively.
Also an absorbed cut-off powerlaw ($\chi_\nu^2=0.97/256$) provides an acceptable fit, with \mbox{$N_{\mathrm{H}}=(1.1\pm0.3)\times10^{21}$\,cm$^{-2}$}, 
photon-index \mbox{$\Gamma=1.1\pm0.2$}, high-energy cut-off \mbox{$E_{\mathrm{high}} = 5^{+2}_{-1}$}, and an observed 0.3--10 keV flux of \mbox{$(6.1\pm0.3)\times10^{-13}$\,\flux}. However, since the disc models appear to provide slightly better results, in the following we will only consider fits based on diskbb and diskpbb models. 

The hardness ratio in Fig.\,\ref{fig:XMM_folded} (bottom panel) clearly shows that the variability in \src \ is associated with marked spectral variations, with spectra becoming harder any time the flux peaks. Therefore, the above values have to be intended only as rough estimates that reflect the general (and approximate) spectral properties of \src . Phase-resolved spectroscopy doubtlessly constitutes a more suitable approach to probe the spectral properties of our target. 
Hence we analysed the EPIC-pn and MOS spectra of \src \ in four phase ranges, labelled as {\it low}, {\it shoulder}, {\it rise} and {\it decay} (see Fig.\,\ref{fig:XMM_folded}, top panel). 

\begin{figure*}[ht]
\begin{center}
\includegraphics[width=0.46\textwidth]{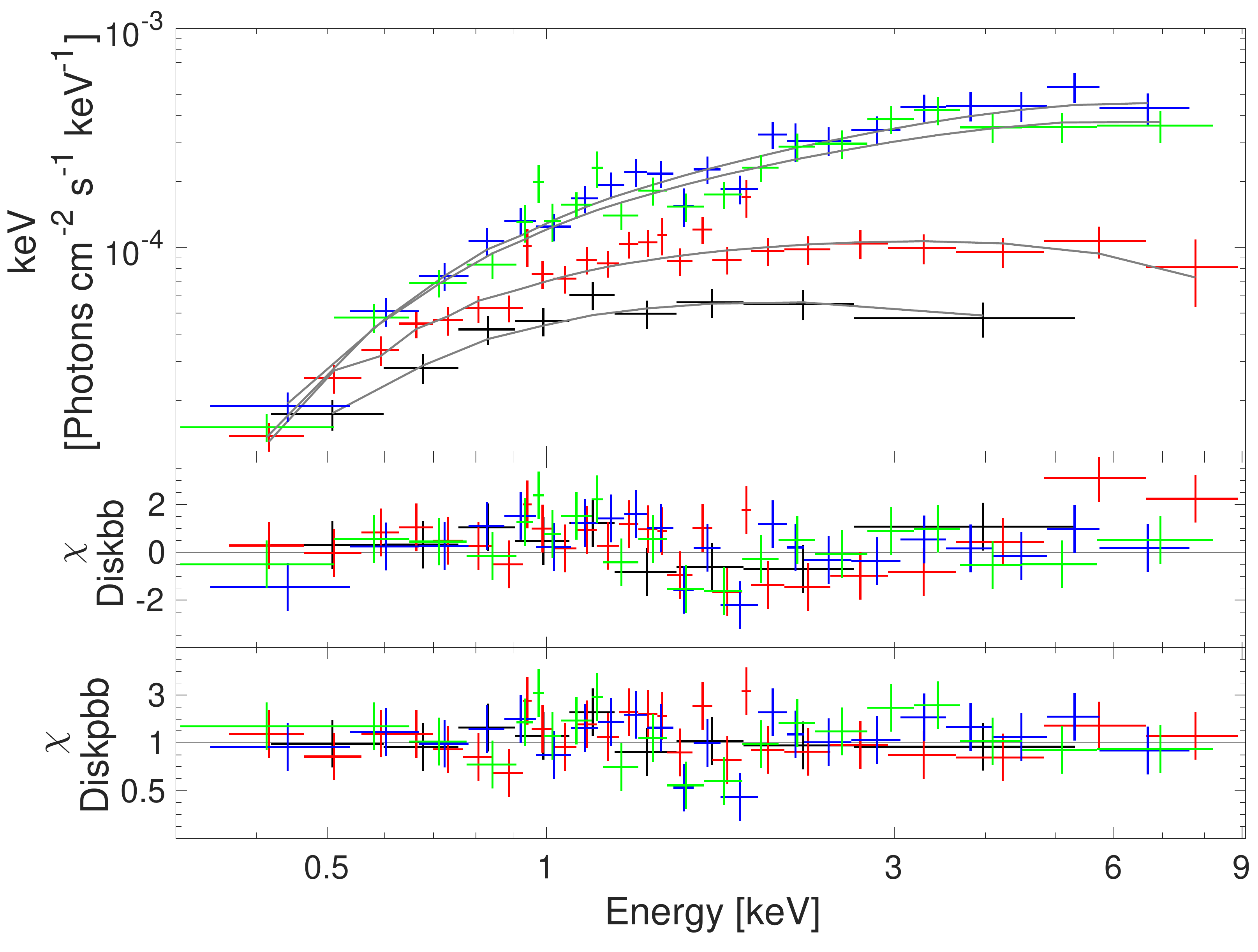}
\includegraphics[width=0.5\textwidth]{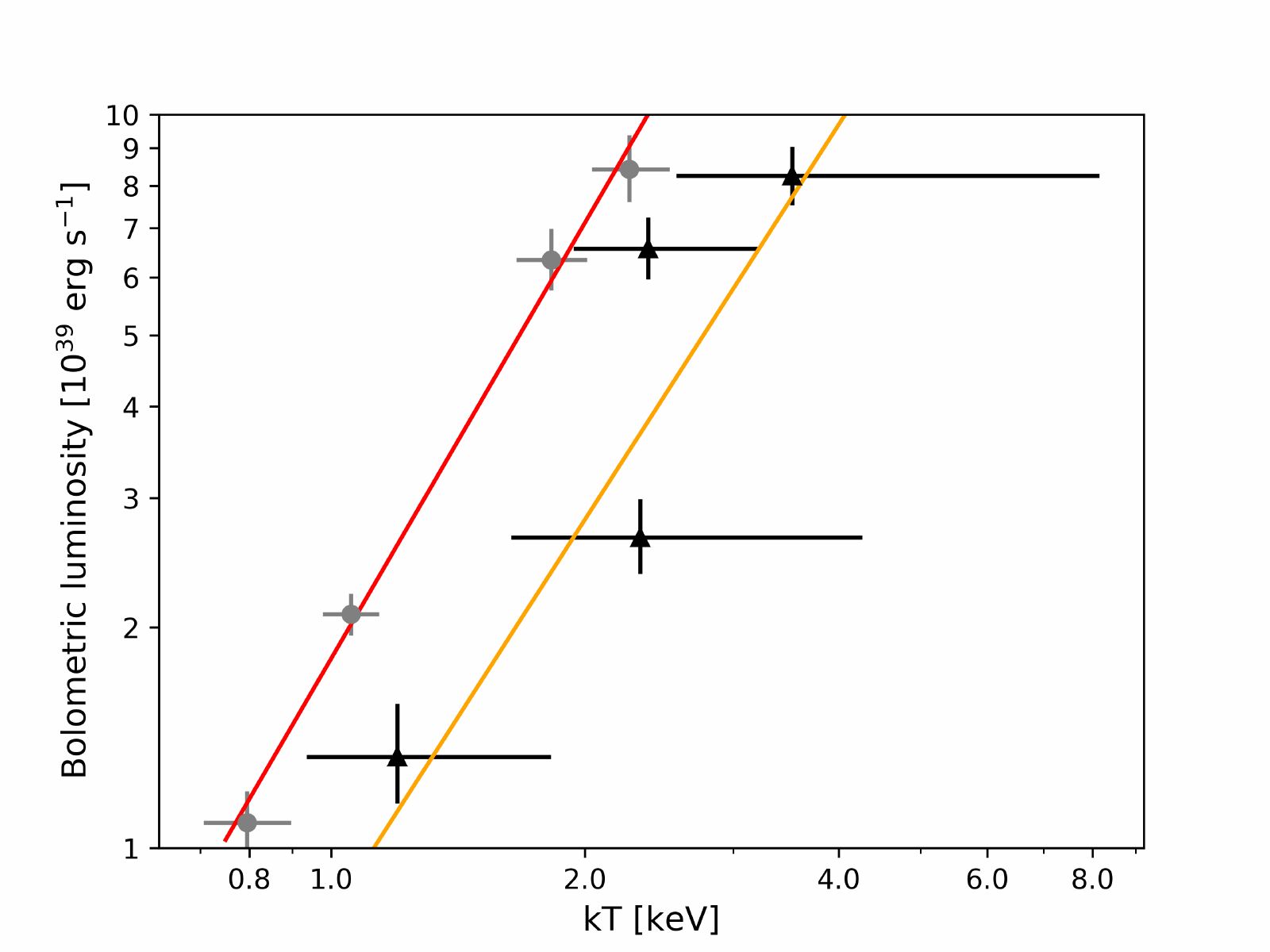}
\caption{Left: unfolded \xmm\ spectra of the phase resolved epochs fitted with a  diskbb model. Colors mark different phases: black is \textit{low}, red is \textit{shoulder}, blue is \textit{rise} and green is \textit{decay}. Data were rebinned for displaying purposes only. For clarity, we show only the EPIC-PN spectra. Right: disc luminosity as a function of the temperature for the  diskbb (grey points) and diskpbb (black triangles) model fitted with $L\propto T^{1.97\pm0.01}$ (red solid line) and $L\propto T^{1.8\pm0.7}$ (yellow solid line), respectively.}
\label{fig:phase_res_spec}
\end{center}
\end{figure*}

As for the average spectrum, phenomenological models such as an absorbed  powerlaw, a  diskbb, or a  diskpbb, all provide adequate fits to the individual phase intervals spectra, and return significantly varying flux and parameters.
Fitting simultaneously the spectra 
with a diskbb, 
leaving free to vary both the disc temperature and its normalization for each phase range while keeping a common column density, provides an acceptable fit ($\chi^2/\mathrm{dof}=1.10/221$; Table\,\ref{tab:phase_spec}), which however retains some structured residuals to the data (see Fig.\,\ref{fig:phase_res_spec}, middle panel). According to this model, the disc unabsorbed bolometric luminosity correlates with the inner disc temperature, following a $L\sim T^{(1.97\pm0.01)}$ relation (see Fig.\,\ref{fig:phase_res_spec}, right panel). Interestingly, a relation of the form $L\sim T^2$ is expected for a slim disc. The deviations of a slim disc from a standard disk-blackbody might thus be at the origin of the structured residuals that we observe. 

Based on the above consideration, we  substituted the  diskbb with a  diskpbb model, where the radial dependence of the disc temperature has a powerlaw profile of the form $r^{-p}$, with $p = 3/4$ for a standard disc and $p = 1/2$ for a slim disc. When fitting this model to the data, we linked the $p$ parameter across the two spectra at higher flux, and across the two spectra at lower flux, respectively. This configuration returns a good description of the spectra, slightly preferred by the data with respect to the previous model   ($\chi^2/\mathrm{dof}=0.96/219$, see Table\,\ref{tab:phase_spec} and Fig.\,\ref{fig:phase_res_spec}, top panel for the best-fit and bottom panel for the model residuals), with a $p$-parameter of $0.63\pm0.04$ and $0.53\pm0.04$ for the high and the low flux spectra, respectively. 
The disc temperature for the four phase range lies between 1 and 4\,keV, although the values are less constrained than in the previous fits.
Using this model, we obtain a luminosity-temperature relation of the form $L\sim T^{(1.8\pm0.7)}$, which is fully consistent with the relation expected in the presence of a slim disc. 
A rough estimate of the range of the inner radii from the  diskpbb model normalisation lies at $\approx$15--40\,km, adopting an inclination angle of 60\textdegree\ and with no colour correction applied. These values must be taken as merely indicative, as they clearly do not reflect the actual disc truncation radii, but rather suggest that the truncation radius may change with varying flux.
The simplified models we adopted may not fully describe the complex spectral variations observed in \src, but the limited signal-to-noise ratio and the lack of data at high energies of our spectra does not allow us to test more sophisticated models. 

Finally, we considered the typical spectral modelling employed to explain the ULX phenomenology, which explicitly assumes super-Eddington accretion around a stellar mass compact object (either a BH or a neutron star, NS). In such a scenario, at least two main spectral thermal components are identified: a cold one that is usually associated with extended and optically thick winds (launched by the super-Eddington disc), and a hot one ascribed to an advection-dominated disc \citep[e.g.][]{middleton15}.
We attempted to model the spectra of \src\ with a cold blackbody plus a hot  diskbb component\footnote{A  diskpbb component would be more adequate for such a scenario; however, the data quality does not allow to put firm constraints on the parameters of this model.}. Since limited variability is expected  from the wind due to its large radial extension \citep{middleton15}, we linked both the column density and the blackbody parameters across the four phases, while we left the  diskbb parameters free to vary. 
This model provides a good fit of the data ($\chi^2/\mathrm{dof}=0.9/220$), which gives a column density of $(1.0\pm0.3)\times10^{21}$\,cm$^{-2}$ and cold blackbody temperature of $0.24\pm0.04$\,keV, with an emitting radius of $1000\pm350$\,km. The hot  diskbb components gives a common temperature of $1.4\pm0.2$\,keV for the {\it low} and {\it shoulder} spectra, while for the {\it rise} and {\it decay} we obtained a temperature of $2.5_{-0.3}^{+0.4}$\,keV and $2.0_{-0.2}^{+0.3}$\,keV, respectively. Based on the  diskbb fits, assuming no color correction factor and a disk inclination of 60\textdegree, the inner radii are $40_{-11}^{+14}$\,km, $60_{-15}^{+18}$\,km, $42_{-9}^{+9}$\,km, and $56_{-10}^{+12}$\,km, for the {\it low}, {\it shoulder},  {\it rise} and  {\it decay} spectra, respectively.

\section{Other X-ray observations}\label{sect:otherdata}

The field of \src\ was observed with \cxo, \swift, and \nus\ (largely overlapping with the \xmm\ observation; see Table\,\ref{table:obs}).

In 2008, \cxo\ caught the source at a rather low luminosity. 
Since the \cxo\ observation does not provide enough counts for a meaningful spectral analysis, in order to estimate the luminosity from the count rate we arbitrarily assumed that the spectrum is described by a blackbody with temperature of 0.4\,keV, which is the temperature we obtain by fitting a blackbody model to the $low$ spectrum.  We obtain a luminosity of $\approx 4\times10^{37}$\,\lum\ (consistent values are obtained with other plausible models, for example $\approx$$6\times10^{37}$\,\lum\ adopting a power-law model with slope $\Gamma=2$).

\src \ was never convincingly detected in the \swift\ exposures, due to the presence of contaminating sources very close to the target's fiducial position in the \swift \ images, as well as of diffuse X-ray emission from NGC\,3261.
The deepest \swift\ upper limits are of the order of  $8\times10^{38}$\,\lum\ ($10^{-3}$\,XRT counts s$^{-1}$ correspond to $\approx 1.3\times10^{38}$\,\lum).

In the \nus \ data \src\ is detected with a significance exceeding 5$\sigma$ up to 15--20 keV. We extracted the background-subtracted 3--15 keV light-curve, combining the data from the \nus/FMPA and \nus/FMPB instruments. The light curve shows flares similar to those observed in the \xmm \ data, which appear to occur at the same time when \nus \ and \xmm \ were observing at the same time. 
However, due to the low signal-to-noise ratio of the data, the contamination from a nearby ($\sim$30$''$) bright source, and the orbit gaps, we refrained from using the \nus\ data for the spectral or timing analysis.

\section{Discussion}

The high Galactic latitude of this source and the faint optical counterparts (see Appendix\,\ref{sec:HST}) make an association of \src\ with a foreground X-ray source (such as a CV or a polar) quite unlikely. We thus approach \src \ as a new ULX, which enriches the still small population of known transients within this class.

The most striking feature of \src\ is the similarity of its light-curve with that of the Galactic BH binary GRS\,1915+105 in one of its many variability states: the $\rho$ class variability \citep{Belloni2000}, sometimes affectionately referred to as the \textit{heartbeats},  which is characterised by a quasi-periodic modulation in the flux.
Such marked short time-scale variability is clearly at odds with most of the ULXs, where only a small sample of sources is highly variable on time-scales shorter than a few hours \citep[e.g.][]{Earnshaw2017,Pintore2020}. \src \ can be classified as {\it broadened-disc} or {\it hard-ultraluminous} source \citep{sutton13}, and, by showing both relatively hard spectra and obvious variability, may depart from the common variable-when-softer connection observed in other ULXs (\citealt{Sutton2013, Middleton2015}). 

Aside form \GRS, a few more objects at times show a $\rho$-like variability: IGR\,J17091--3624, MXB\,1730--335, and the AGN in GSN\,069. 
IGR\,J17091--3624 is thought to be powered by a stellar mass BH, and is considered a fainter relative of GRS\,1915+105, with which it shares many characteristic variability patterns  (\citealt{Altamirano2011}). 
MXB\,1730--335, also known as the Rapid Burster, has been recently shown to display---among many other variability patterns---two of the many characteristic variability classes observed in GRS\,1915+105 (the $\rho$ and $\theta$ class, see \citealt{Bagnoli2015}), despite hosting a confirmed NS (\citealt{Hoffman1978}). GSN\,069, which is powered by a super-massive BH with a mass of the order of $\sim 10^5 M_{\odot}$, recently started to show a flux modulation reminiscent of the heartbeats \citep{Miniutti2019}. Such a modulation, unsurprisingly, occurs with a much longer periodicity than in \GRS \ (approximately 9 hours), ascribed to the large mass of this object. 


When compared in some detail, the light-curves of \src, \GRS, \IGR\ and the \RB, all show the same structure, constituted by a low-flux plateau, a shoulder, and a main peak, the latter typically asymmetric, with a slower rise to the peak (in our case equivalent to the sum of the time interval covering the \textit{shoulder} and $T_{\rm rise}$) and a fast decay. In this matter, \GSN\ represents an exception, with its very symmetric peaks and non-detectable pre-peak shoulder. 
What clearly differentiates these qualitatively similar light-curves is the recurrence time of the main peaks. From fastest to slowest, the recurrence times are $\sim$5--100\,s (\IGR), 40--1000\,s (\GRS), $\sim$350--450\,s (\RB), $\sim$3200--3700\,s (\src), and $\approx$32.5\,ks (\GSN). 

The soft X-rays spectrum (below 10\,keV) at the time of the heartbeat in \src \ appears to be thermal, with the emission well-modelled by a disk-like component, and relatively hard. Our best fits to the phase-resolved spectra of \src \ indicate that both the disk temperature and the inner disc truncation radius vary across different phases, but while the former clearly increases with flux, our data do not allow us to make a conclusive statement on the evolution of the inner disc radius. 
Similar properties have been observed in all the aforementioned sources (which---we remind the reader---include a NS and a BH Galactic X-ray binary, as well as a super-massive BH), with the addition that in all these sources the inner disc radius consistently increases with flux during the heartbeats. 
In particular, in \GRS---the first source to have shown this particular variability state---the above spectral behavior has been interpreted as the result of a limit-cycle instability driven by the Lightman–-Eardley radiation pressure instability (\citealt{Lightman1974}, \citealt{Belloni2000}). An excess density wave propagates in the innermost portion of the accretion disc, eventually inducing local Eddington accretion that causes the flux to peak, while the accretion disc reacts by increasing its truncation radius at an almost constant temperature. Our results are consistent with this picture, although the limited statistical quality of our data does not allow us to make any stronger statement. The radiation pressure limit-cycle instability has been invoked to explain the heartbeats also in \IGR\ (\citealt{Altamirano2011}), \GSN\ (\citealt{Miniutti2019}), and even in the \RB, where the instability might develop in a portion of the accretion flow unaffected by type-I X-ray bursts eventually emitted from the neutron star surface (see \citealt{Maselli2018}). 

The fits to the phase-resolved spectral return a relation between bolometric luminosity versus disc temperature of the form $L\propto T^p$, where $p\approx$2, 
which indicates that the accretion disc around \src \ is likely a slim disk. A similar powerlaw relation has been found by \citealt{Miniutti2019} for \GSN , who found a luminosity-temperature relation with index $p = 4$. For what concerns \GRS \ and \IGR, a comparison with \src \ is much more difficult: possibly due to the much better signal-to-noise ratio of the data from these sources, and/or a much larger amount of data available, it appears that in either sources there is no interval longer than a few seconds over which $L\propto T^4$ (or any other power-law relation with  constant index). Rather, both sources trace hysteresis cycles in the luminosity versus temperature plane with p varying in the 2 to 4 range (but in opposite directions), reflecting a complicated accretion scenario \citep{Altamirano2011,Court2017,Maselli2018}. 
In the case of the \RB, it is unclear whether a correlation between disc temperature and luminosity exists in the first place, as the disc temperature during the heartbeats seems to remain roughly constant in time (\citealt{Bagnoli2015}).  

Among the sources showing heartbeats considered here, both a mass and a distance estimate are available for GRS\,1915+105 ($M \approx 12\,M_{\odot}$, $d \approx 8.6$\,kpc; \citealt{Reid2014}), for the \RB \ ($M \approx 1.5\,M_{\odot}$, $d \approx 8$\,kpc; \citealt{Lewin1976}), and for \GSN \ ($M \approx 4 \times  10^{5}\,M_{sun}$, $z \approx 0.018$; \citealt{Miniutti2019}). This made it possible to establish that the heartbeats occur at $\sim$80--90\% of the source Eddington luminosity ($L_{\mathrm{Edd}}$) in \GRS , between 40 and 90\% $L_{\mathrm{Edd}}$ in \GSN , and only at $\approx$20\% in the \RB \ \citep{Neilsen2011,Miniutti2019,Bagnoli2015}. On the one hand, this suggests that about Eddington accretion rates might not be a necessary condition for the heartbeats to occur, a fact already noted by many authors and also supported by hydrodynamical simulations (see e.g. \citealt{Janiuk2015}). On the other hand, the above observation implies that we cannot easily assume that \src \ is accreting at close-to-Eddington rates by comparison with, e.g., \GRS. This fact becomes even more striking when we consider that while \GRS , \IGR, the \RB , and \GSN \  undoubtedly show similar heartbeats light-curves and spectral properties, all these sources (and especially the three stellar-mass systems) also show a large variety of widely unexplained additional variability states. Therefore, without more observations, better signal-to-noise, and more in-depth studies, we are not in a position to exclude that all the similarities that we have identified among the above sources are purely coincidental.

\smallskip

We are now left with the daunting task of discussing the possible nature of the compact object hosted in \src . 
Having excluded the obvious presence of type-I X-ray bursts in our X-ray data by visual inspection of the data, we performed an accelerated search for coherent signals in the \xmm \ observation, which did not reveal any significant feature (see  Appendix\,\ref{sec:pulses} for details on the analysis). Therefore we have no direct proof of the presence of a surface, which would unquestionably indicate the presence of a NS.
Our spectral analysis does not offer any insight into this matter either,  as no clear similarity or difference between \src \ and either the \RB \ (hosting a confirmed NS) or any of the other systems we considered (all hosting confirmed or candidate BHs) has emerged. 
Our only option is to attempt a mass estimate of the compact object based on the information we have on \src , aware of the fact that the mass of the accreting object might not the dominant factor that determines its variability time-scale (see, e.g., \citealt{Massaro2020a, Massaro2020b}, for a case study on \GRS). We identified three different ways to do this, all depending on fairly strong assumptions, and thus limited by important caveats.

Firstly, by assuming that the maximum luminosity reached by \src \ equals its Eddington limit (which is not necessarily a sensible assumption, as discussed above) based on our spectral analysis we obtain a mass of approximately 75\,$M_{\odot}$. By assuming that  \src \ is accreting at 10 times the Eddington accretion rate instead---which is not unusual in NS powered ULXs (see, e.g.,  \citealt{israel17})---this mass moves down to 7.5\,$M_{\odot}$, a value comfortably consistent with the average expected mass of BHs in Galactic binaries systems (\citealt{Ozel2010}).
Secondly, we observe that all the most relevant timescales in accreting systems (e.g., the viscous, thermal, and perturbation propagation time-scales) are proportional to the dynamical time scale, $t_{\mathrm{dyn}} \approx \frac{GM}{c^3} r^{3/2}$, where $r$ is expressed in units of the gravitational radius $R_{\mathrm{g}} = GM/c^2$. 
If we make the (strong) assumptions that the heartbeats arise from the same portion of the accretion flow in two sources, and that the two accretion flows have the same characteristics (e.g. same disc scale height $H/R$ and viscosity parameter $\alpha$), we obtain that any characteristic time scale simply scales linearly with the mass of the object.\footnote{This is because both the radius and the accretion flow parameters $H/R$ and $\alpha$ can be eliminated from the equations when comparing timescales, because they are assumed to be constant across the sources considered.} Taking $M = 10\,M_{\odot}$ for \GRS, and a recurrence time of 100\,s and 3500\,s, respectively, for \GRS\ and \src, it follows that \src \ should have a mass of the order $M \approx 350\,M_{\odot}$. 
It is immediately clear, however, that such a rough estimate is strongly sensitive to the choice of (i) the characteristic time-scale used, which can vary significantly even for one single object (up to 3 orders of magnitude for, e.g., \IGR \ and \GRS); and (ii) the object used for the comparison (e.g., \GRS \ as opposed to the \RB \ or \GSN). This indicates that our assumptions are most likely too simplistic, and the above mass estimate can easily vary by orders of magnitude depending on our relatively arbitrary choices. 
Finally, we can attempt a mass estimate by assuming that the heartbeats recurrence times are governed by the radiation-pressure limit-cycle instability. We followed the approach adopted by \cite{Miniutti2019} and we used an empirical relation derived from numerical simulations of AGN that links the BH mass, the instability recurrence time, and the ratio $A$ between the minimum and maximum luminosity sampled during the instability duty-cycle---that is, \mbox{$M_{\mathrm{BH}} = 0.45 T_{\mathrm{rec}}^{0.87}$ $A^{-0.72} (\alpha/0.02)^{1.88}$} (\citealt{Janiuk2004})---to compare once more \src \ and \GRS . By making the same assumptions as above on the accretion flow, and taking $A = 0.3$ for \GRS\ (based on \citealt{Neilsen2011}), and $A = 0.13$ from this work for \src, we find that the latter should have a mass of the order $M \approx 400\,M_{\odot}$. 
As mentioned above, it is unlikely that $\alpha$, $H/R$, and the size of the portion of the accretion flow generating the heartbeats are the same for \src \ and \GRS \ (or any other system that might be used for comparison). It follows that, once more, the above mass estimate is not really constraining, as it strongly depends on our assumptions. 

Based on all the above considerations, we cannot make a strong statement on the nature of the compact object in \src, which might very well be a stellar-mass BH as well as a NS, or even a more exotic intermediate-mass BH. We believe the latter possibility less likely, as the existence of an intermediate-mass BH would be difficult to explain in terms of galactic evolution and dynamics in a galaxy powered by a relatively light super-massive BH, and lacking any clear sign of interaction with other galaxies. 

Clearly, the marked variability observed in \src \ may be due to a number of processes different from the radiation-pressure  limit-cycle  instability, and not necessarily instability-driven (see, e.g., \citealt{middleton18} for a number of possibilities involving precession, and  \citealt{Bagnoli2015b} for a list of magnetic models). However, the fact that we are currently unable to constrain the mass of the accreting object in this source prevents us from making any meaningful comparison between its observed variability and that predicted by essentially any existing model. New observations and more data are needed in order to clarify what physical processes might underlie the observed properties of \src .



\smallskip

\acknowledgments
\noindent We thank the anonymous referee for their helpful comments, which contributed to improve this work.
This research has made use of data produced by the EXTraS project, funded by the European Union's Seventh Framework Programme under grant agreement n.\,607452.
The scientific results reported in this article are based on observations obtained with \xmm, an ESA science mission with instruments and contributions directly funded by ESA Member States and NASA.
This work was supported by the Oxford Centre for Astrophysical Surveys, which is funded through generous support from the Hintze Family Charitable Foundation.
We acknowledge funding in the framework of the ASI--INAF contract \mbox{n.\,2017-14-H.0} and of the PRIN MIUR 2017 ``UnIAM (Unifying Isolated and Accreting Magnetars)''.
We acknowledge the INAF computing centre of Osservatorio Astrofisico di Catania for the availability of computing resources and support under the coordination of the CHIPP project.

\newpage 

\begin{table*}[hbt!]
\centering       
\caption{Journal of the X-ray observations of \src. The count rates are in the 0.3--10 keV energy range, except for \nus, for which we used the 3--15\,keV energy range. Uncertainties are at 1$\sigma$ confidence level, upper limits are at 3$\sigma$.
}
\begin{tabular}{ccccc}
\hline      
Instrument & Obs.ID & Date & Exposure & Count rate\\ 
 & & & (ks) & (Counts s$^{-1}$) \\
\hline   
\hline
\cxo/ACIS-S & 9278 & 2008-03-06 & 21.2 & $(7.5\pm2.7)\times10^{-4}$  \\
\swift/XRT & 00045607001 & 2011-08-01 & 0.9 & $<$$1.3\times10^{-2}$ \\
\swift/XRT & 00045607002 & 2011-08-02 & 3.8 & $<$$8\times10^{-3}$ \\
\swift/XRT & 00045607003 & 2011-08-04 & 2.0 & $<$$1.1\times10^{-2}$ \\
\swift/XRT & 00045607004 & 2011-08-05 & 2.5 & $<$$9\times10^{-3}$ \\
\swift/XRT & 00045607005 & 2011-08-08 & 0.3 & $<$$4.2\times10^{-2}$\\
\swift/XRT & 00045607006 & 2011-08-10 & 0.4 & $<$$3.8\times10^{-2}$\\
\swift/XRT & 00045607007 & 2011-08-14 & 3.9 & $<$$9\times10^{-3}$ \\
\swift/XRT & 00045607008 & 2012-08-12 & 4.6 & $<$$6\times10^{-3}$ \\
\swift/XRT & 00045607009 & 2012-08-14 & 6.0 & $<$$6\times10^{-3}$ \\
\swift/XRT & 00045607010 & 2012-08-15 & 3.3 & $<$$8\times10^{-3}$ \\
\swift/XRT & 00045607011 & 2012-08-17 & 2.9 & $<$$1.2\times10^{-2}$ \\
\swift/XRT & 07006864001 & 2017-02-28 & 0.1 & $<$0.28 \\
\swift/XRT & 07006862001 & 2017-03-10 & 0.1 & $<$0.27\\
\swift/XRT & 07006864002 & 2017-03-15 & 0.5 & $<$$2.6\times10^{-2}$ \\
\swift/XRT & 00088212001 & 2017-12-15 & 1.1 & $<$$5.4\times10^{-2}$ \\
\nus/FPM\,A+B & 60371002002 & 2017-12-15 & 30.8 & $(4.5\pm0.4)\times10^{-3}$\\
\xmm/pn & 0795660101 & 2017-12-16 & 33.7 & $0.135\pm0.002$\\
\xmm/MOS1 & 0795660101 & 2017-12-16 & 34.9 & $0.041\pm0.001$\\
\xmm/MOS2 & 0795660101 & 2017-12-16 & 34.9 & $0.040\pm0.001$\\
\swift/XRT & 00088212002 & 2017-12-20 & 0.9 & $<$$1.3\times10^{-2}$ \\
\hline                  
\end{tabular}\label{table:obs} 
\end{table*}

\begin{table*}
\centering
\caption{Properties of the QPOs detected in the EPIC-pn and MOS data from 4XMM\,J111816.0--324910. The \textit{main peak} is the most prominent peak in the PDS from both the EPIC-pn and MOS data. The \textit{Secondary peak 1 and 2} are the two peaks visible on the right-hand side of the main peak. The \textit{left shoulder} is the low-amplitude feature visible at the left-hand side of the main peak, and only detected in the EPIC-pn data. Uncertainties are give at a 1 $\sigma$ level.}
    \begin{tabular}{lcccc}
    \hline
    \textbf{Instrument}  &\textbf{Component} & \textbf{Frequency [Hz]} & \textbf{$Q$ factor} & \textbf{Significance [$\sigma$]} \\ 
    \hline
    \hline
            & Main peak        & (3.022 $\pm$ 0.003)$\times 10^{-4}$           &   17.9       &    11.9      \\
    EPIC-pn & Left shoulder    & (2.22$^{+0.01}_{-0.01}$)$\times 10^{-4}$   &   unresolved &    5.4       \\
            & Secondary peak 1 &  (5.2 $\pm$ 0.1)$\times 10^{-4}$            &   unresolved &    6.8       \\
            & Secondary peak 2 &  (7$\pm$ 2)$\times 10^{-4}$                   &   unresolved &    5.9       \\    
    \hline 
    \end{tabular}
    \label{tab:QPOs}
\end{table*}

\begin{table*}
\renewcommand{\arraystretch}{1.4}
\centering
\begin{tabular}{c|l|llll|c}
\hline
Model & Parameter & {\it low} & {\it shoulder} & {\it rise} & {\it decay} & $\chi_{\nu}^2/\text{dof}$ \\
\hline
\hline

\multirow{4}{*}{ diskbb} & N$_{\rm H}$ [$10^{20}$ cm$^{-2}$] & \multicolumn{4}{c|}{$4_{-1}^{+2}$ } & \multirow{4}{*}{1.11/221}  \\
 & kT [keV] & $0.80_{-0.09}^{+0.1}$ & $1.06_{-0.08}^{+0.08}$ & $2.3_{-0.2}^{+0.3}$ & $1.8_{-0.2}^{+0.2}$ & \\
 & Norm. [$10^{-3}$] & $23_{-8}^{+13}$ & $14_{-3}^{+5}$ & $2.8_{-0.8}^{+1}$ & $4.9_{-1}^{+2}$ & \\
 & L [$10^{39}$]$^a$ & $1.0_{-0.1}^{+0.1}$ & $2.0_{-0.1}^{+0.1}$ & $7.4_{-0.6}^{+0.6}$ & $5.9_{-0.5}^{+0.5}$ & \\
\hline

\multirow{5}{*}{diskpbb} & N$_{\rm H}$ [$10^{20}$ cm$^{-2}$] &  \multicolumn{4}{c|}{$15_{-4}^{+4}$ } & \multirow{5}{*}{0.96/219}  \\
 & kT [keV] & $1.2_{-0.3}^{+0.6}$ & $2.3_{-0.7}^{+1.9}$ & $3.5_{-0.9}^{+4.6}$ & $2.4_{-0.4}^{+0.8}$ & \\
 & $p$ & \multicolumn{2}{c}{$0.53_{-0.03}^{+0.04}$} &  \multicolumn{2}{c|}{$0.63_{-0.03}^{+0.04}$} & \\
 & Norm. [$10^{-4}$] & $16_{-10}^{+30}$ & $2_{-2}^{+9}$ & $3_{-3}^{+9}$ & $11_{-8}^{+20}$ & \\
 & L [$10^{39}$]$^a$ & $1.3_{-0.2}^{+0.2}$ & $2.7_{-0.3}^{+0.3}$ & $8.3_{-0.7}^{+0.8}$ & $6.6_{-0.6}^{+0.7}$ & \\
\hline

\multirow{7}{*}{bbody+diskbb} & N$_{\rm H}$ [$10^{20}$ cm$^{-2}$] &  \multicolumn{4}{c|}{$10_{-3}^{+3}$ } & \multirow{7}{*}{0.93/220}  \\
 & kT$_{\text{bbody}}$ [keV] & \multicolumn{4}{c|}{$0.24_{-0.03}^{+0.04}$ } & \\
 & Norm. [$10^{-7}$] & \multicolumn{4}{c|}{$10_{-8}^{+12}$ } & \\
 & L$_{\text{bbody}}$ [$10^{38}$]$^a$ & \multicolumn{4}{c|}{$4.1_{-0.9}^{+0.9}$ } & \\
& kT$_{\text{diskbb}}$ [keV] & \multicolumn{2}{c}{$1.4_{-0.2}^{+0.2}$ } &  $2.5_{-0.3}^{+0.4}$ & $2.0_{-0.2}^{+0.3}$ & \\
 & Norm. [$10^{-3}$] & $1.8_{-0.9}^{+1}$ & $4_{-2}^{+3}$ & $1.9_{-0.7}^{+1}$ & $4_{-1}^{+2}$ & \\
 & L$_{\text{diskbb}}$ [$10^{39}$]$^a$ & $0.8_{-0.2}^{+0.1}$ & $1.8_{-0.2}^{+0.2}$ & $7.5_{-0.6}^{+0.6}$ & $5.6_{-0.5}^{+0.5}$ & \\
\hline
\end{tabular}
\caption{Best-fits of the four phase resolved spectra. Errors are quoted at a $90\%$ confidence level for each parameter of interest.\\
$^a$ Unabsorbed 0.3--10 keV luminosity.}
\label{tab:phase_spec}
\end{table*}


\facilities{\xmm\ (EPIC), \swift\ (XRT), \cxo\ (ACIS), \nus, \emph{HST} (ACS)}
\software{SAS \citep{gabriel04}, FTOOLS \citep{blackburn95}, XSPEC \citep{Arnaud1996}, CIAO \citep{fruscione06}}
 
\bibliographystyle{aasjournal.bst}
\bibliography{biblio}
\clearpage

\appendix
\section{Search for rapid pulsations}\label{sec:pulses}
\begin{figure}[ht]
\centering
\hspace{-8mm}
\includegraphics[width=9cm]{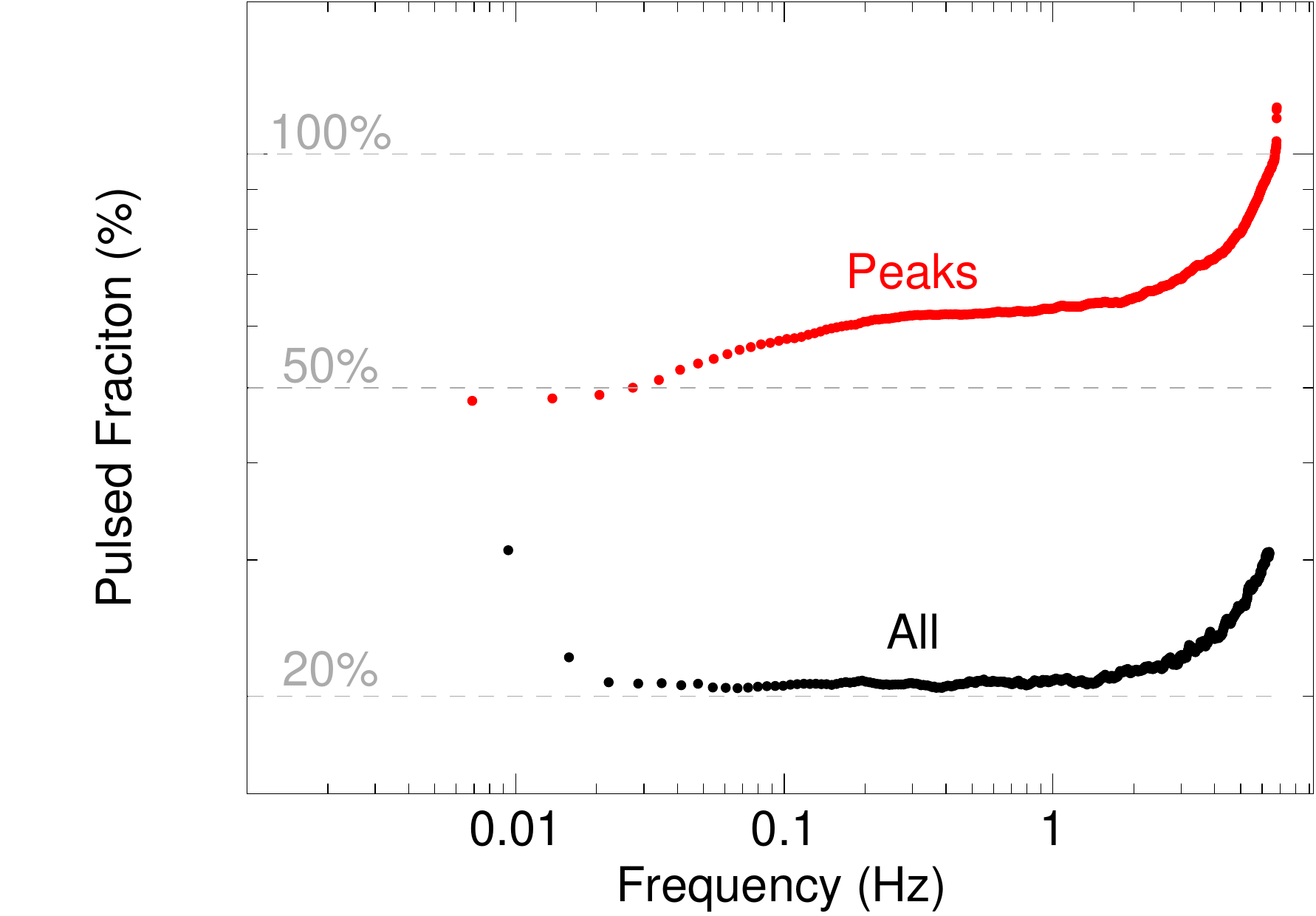}
\caption{\label{figUL} Upper limits on fast  pulsations inferred from an accelerated search taking into account the possible presence of a first period derivative component $\dot{P}$ and for two different cases: ``All" including all the source events and ``Peaks" only for the brightest intervals of the QPOs. In the former case we inferred 20--30\% pulsed fraction at 3$\sigma$ between 150\,ms and 100\,s, and about 50--90\% pulsed fraction for the "Peaks".}
\end{figure}

We considered the possibility of a fast-spinning pulsar in the system \citep[see, e.g.,][]{israel17}.
For the \xmm\ data set, after correcting the times  to  the  Solar system  barycenter, we performed an accelerated search for signals by correcting the time of arrivals (ToA) of each photon in order to account for time shifts corresponding to those originated by a first period derivative component in the range \mbox{$|\dot{P}/P|< 10^{-6}$}\,Hz, and then by looking for peaks above a 3.5$\sigma$ local detection threshold in the corresponding power spectral density (PSD) and taking onto account for the possible presence of non-Poissonian noise components \citep{israel96}. We performed the search both over the whole sample of available ToA and only for those around the quasi-periodic peaks. The search gave negative results in both cases. No candidate signal was found in the 150\,ms to 100\,s period range, with 3$\sigma$ upper limits in the 20--30\% and 50--90\% range for ``all'' the ToA and for the QPOs ``peaks'', respectively (see Fig.\,\ref{figUL}).  

\section{\emph{Hubble Space Telescope} observations}\label{sec:HST}
NGC3621 was observed with the \emph{HST}/ACS instrument in the F435W (B) filter, F555W (V) filter and F814W (I) filter. We retrieved calibrated, geometrically-corrected images from the \emph{Hubble} Legacy Archive (HLA2). We refined astrometry using a set of sources from the 2MASS survey, with a resulting r.m.s. accuracy better than 50 mas per coordinate. We run a source detection using the SExtractor software and we converted count rates to magnitudes using the photometric calibration provided by the HLA pipeline. 

The brightest source consistent with the \cxo\ error circle is extended, and it is very likely a star cluster in NGC3621.

\begin{figure*}[hbt!]
\centering
\resizebox{\hsize}{!}{\includegraphics{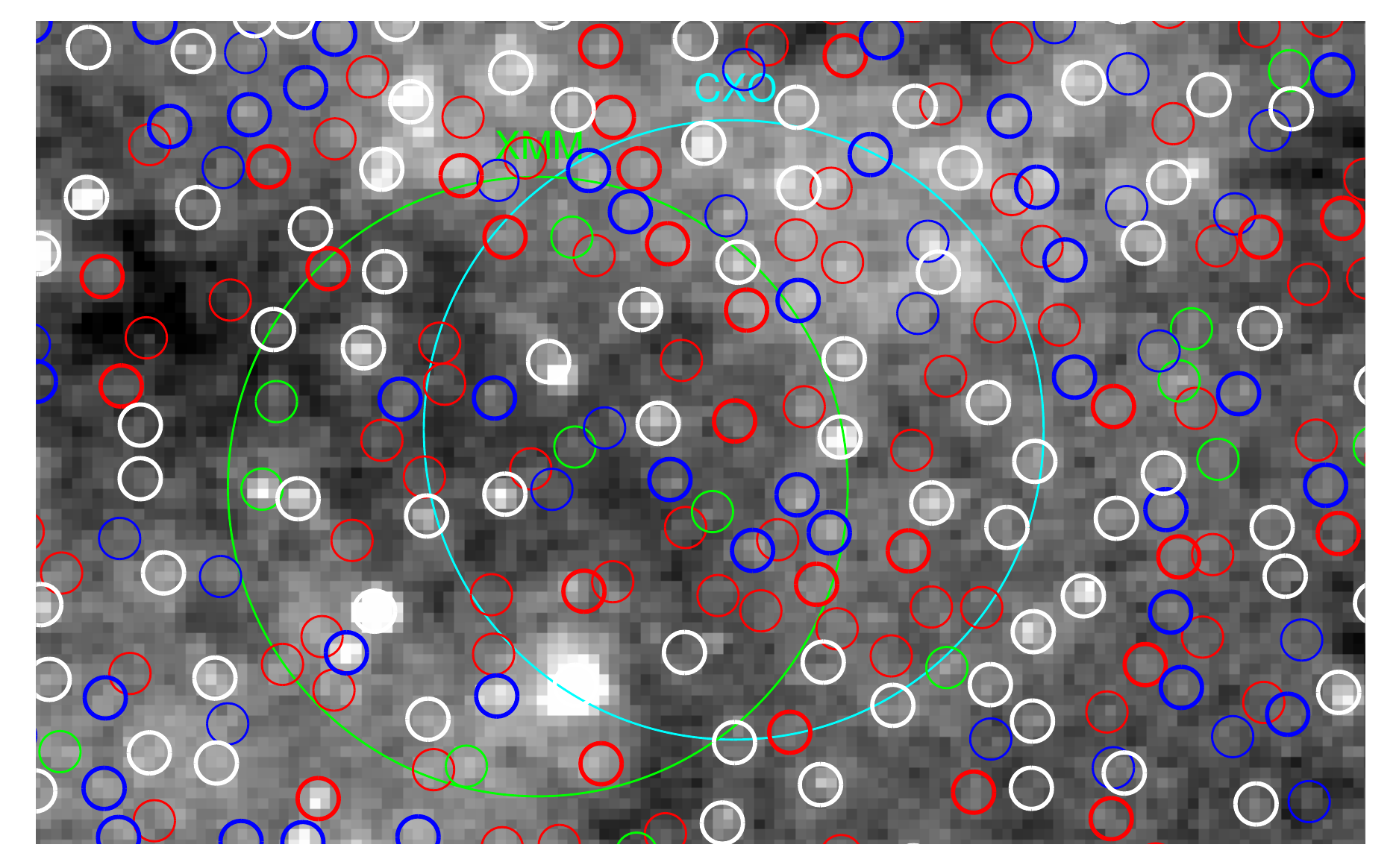}}
\caption{The field as seen by \emph{HST} in the V band, with \xmm\ and \cxo\ error circles superimposed (\cxo: 99\% confidence level; \xmm: 68\% confidence level). White circles: \emph{HST} sources detected in BVI; bold blue: BV, not I; bold red: VI, not B; thin blue: B only; thin green: V only; thin red: I only.  \label{hst:field}}
\end{figure*}

\begin{figure*}[hbt!]
\centering
\resizebox{\hsize}{!}{\includegraphics{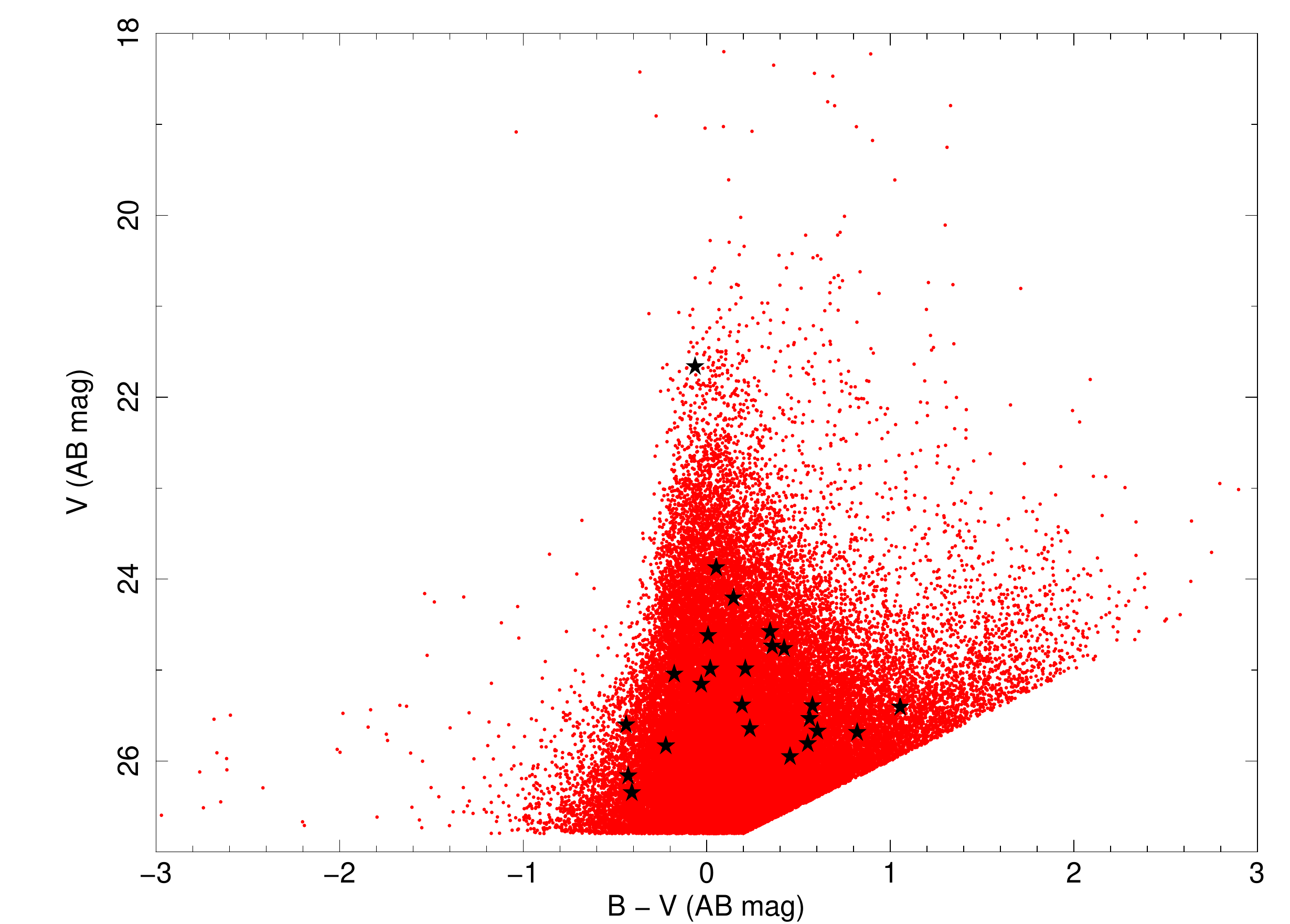}\includegraphics{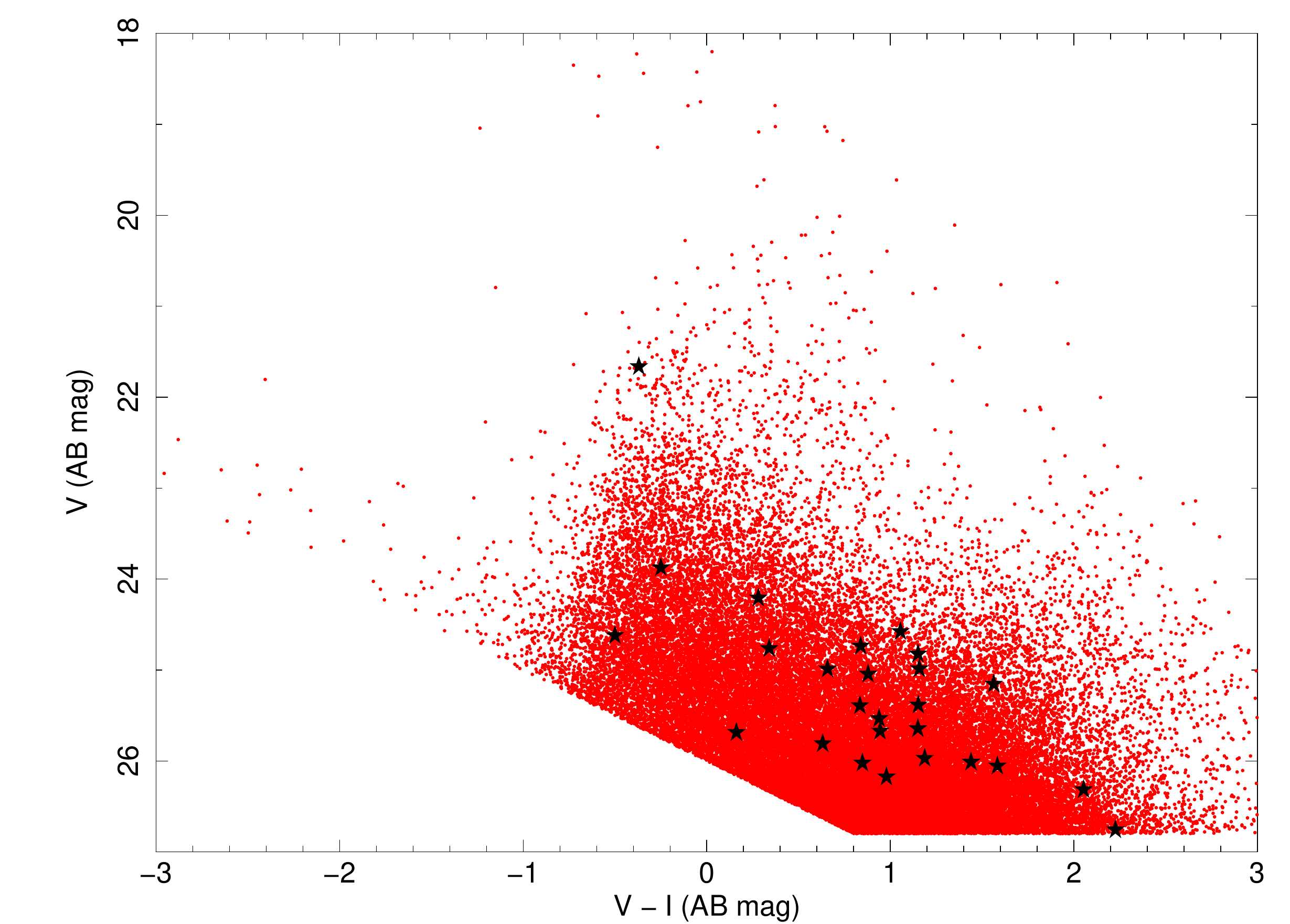}}
\caption{Left: B minus V vs. V color-magnitude diagram, based on sources detected in both filters. Sources lying in the \cxo\ error region are marked with black stars. Right: Same as left figure, but V minus I color-magnitude diagram. \textbf{In both panels we excluded all sources below AB mag 26.8, 27.0 and 26.0 in the B, V and I band, respectively -- such values being a robust estimate of the sensitivity limit of the images, based on the observation of the peak in the number count histogram of magnitudes of detected sources. } \label{hst:colors}}
\end{figure*}


\end{document}